\documentclass[showpacs,twocolumn,superscriptaddress,aps]{revtex4}
\usepackage{epsfig}
\usepackage{graphicx}
\usepackage{bm}

\begin{document}

\title{Fully Quantum Approach to Optomechanical Entanglement}
\author{Qing Lin}
\affiliation{Institute for Quantum Science and Technology, University of Calgary, 
Alberta
T2N 1N4, Canada}
\affiliation{College of Information Science and Engineering, Huaqiao University, Xiamen
361021, China}
\author{Bing He}
\affiliation{Institute for Quantum Science and Technology, University of Calgary, Alberta
T2N 1N4, Canada}
\affiliation{Department of Physics, University of Arkansas, Fayetteville, AR 72701, USA}
\author{R. Ghobadi}
\affiliation{Institute of Atomic and Subatomic Physics, TU Wien, Stadionallee 2, 1020 Wien, Austria}
\author{Christoph Simon}
\affiliation{Institute for Quantum Science and Technology, University of Calgary, 
Alberta T2N 1N4, Canada}

\begin{abstract}
The radiation pressure induced coupling between an optical cavity field and
a mechanical oscillator can create entanglement between them. In previous
works this entanglement was treated as that of the quantum fluctuations of
the cavity and mechanical modes around their classical mean values. Here we
provide a fully quantum approach to optomechanical entanglement, which goes
beyond the approximation of classical mean motion plus quantum fluctuation,
and applies to arbitrary cavity drive. We illustrate the real-time evolution
of optomechanical entanglement under drive of arbitrary detuning to show the
existence of high, robust and stable entanglement in blue detuned regime,
and highlight the quantum noise effects that can cause entanglement sudden
death and revival.
\end{abstract}

\maketitle

\section{Introduction}

The study of optomechanical systems (OMS) has undergone rapid development
over the recent years \cite{R1,R2,R3}. The quantum level of OMS has been
reached in experiments \cite{ex00,ex01, ex010, ex02,ex020, ex03}.
Entanglement is a particularly striking quantum feature. The coupling of the
cavity field of an OMS to the mechanical oscillator under radiation pressure
can lead to their entanglement. This mesoscopic or macroscopic entanglement
possesses both fundamental interest and potential applications.
\begin{figure}[b!]
\vspace{-0cm} \centering
\epsfig{file=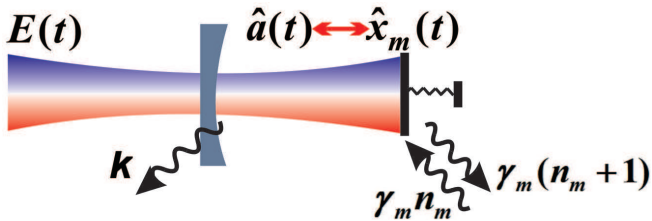,width=1.0\linewidth,clip=} {\vspace{-0.5cm}} \vspace{%
0cm}
\caption{(color online) Setup for coupling the cavity field $\hat{a}(t)$
with the quantum mechanical oscillator. The cavity field build up by the
drive $E(t)$ is entangled with the mechanical oscillator, which is initially
in thermal equilibrium with its environment, via their effective coupling
described by the Hamiltonian $-\protect\sqrt{2}g\hat{x}_m\hat{a}^\dagger\hat{%
a}$. }
\label{Fig:c}
\end{figure}

Theoretically an OMS is often approached via the expansion of its
fluctuations about the mean values of the cavity and mechanical mode
operators, where these mean values are determined by the classical 
equations of motion. This approximation of replacing a quantum system operator with
the sum of a classical value and the accompanying quantum fluctuation has
been widely applied to generic nonlinear quantum systems whose
Heisenberg-Langevin equations are not analytically solvable \cite{semi}.
Most previous studies of optomechanical entanglement (see, e.g. \cite%
{vitali, pater, vitali-07,vitali-08, h-p-08, galve, zou, abdi, G}) concern
that of the fluctuations around the steady state solution of the classical
Langevin equations under continuous-wave (CW) drive. Some other works have
considered the entanglement under periodic \cite{m-e-09, m-e-12, f-g} or
pulsed drive \cite{pulse2}. A common feature of these treatments is that the
linearized dynamics about the fluctuations is based on a specific classical
mean motion as the background, and the entanglement of the fluctuations can
be closely connected to the classical motion of OMS \cite{d-e}. However, the
classical motion of an OMS can be chaotic \cite{mulstability}, so it is not
always possible to quantify this entanglement of fluctuations \cite{d-e}.

Very recently several quantum features of OMS have been studied in
considerable detail. This research includes OMS dynamics under single photon
drive \cite{p1,p2,p3,bhe,p4}, control and generation of OMS quantum states
\cite{s2,s3,s4,s5,s6,s7}, enhancement of OMS nonlinearity for quantum
information processing \cite{n1,n2,n3,n4} and other quantum properties of
OMS \cite{q1,q2}. These studies consider the quantum states associated with
the cavity mode $\hat{a}$ and mechanical mode $\hat{b}$ themselves, as in
Fig. 1, instead of those for their fluctuations. Starting from a separable quantum state of the cavity and mechanical mode,
the optomechanical coupling can entangle them to an entangled quantum state. The less unexplored
entanglement of such fully quantum OMS, which is independent from classical
motion, is the theme to be discussed below. Notice that this type of entanglement was also recently discussed in a different 
approach \cite{et-new}, which works with the numerical simulation based on the approximate Fokker-Planck equation to find the entanglement signature and other properties. 

This paper is organized as follows. In Sec. II, we discuss the dynamics about the OMS in strong drive and weak coupling regime. 
In this regime the quantum states of an OMS keep to be Gaussian. The real-time evolution and quantum noise effect on the entanglement of such Gaussian states are studied with examples in Sec. III.
Then we present a rather detailed discussion about the difference between our concerned entanglement and that of the cavity and mechanical fluctuation in Sec. IV. The conclusions from our study are given in the final section.

\section{dynamics under strong drive and weak coupling}

We consider an OMS driven by a pulsed drive with the central frequency $%
\omega_0$ and arbitrary frequency distribution $E(\omega-\omega_0)$. Its
profile $E(t)e^{i\omega_0 t}$ in time domain is related to $%
E(\omega-\omega_0)$ by the Fourier transform. The drive reduces to a CW one
when $E(t)$ is constant. Without cavity and mechanical damping, one has the
unitary evolution operator $U(t,0)=\exp\{-iH_0t\}\mathcal{T}\exp\{-i\int_0^t
d\tau H_S(\tau)\}$ for the OMS, where $H_0=\omega_c \hat{a}^{\dagger}\hat{a}%
+\omega_m \hat{b}^{\dagger}\hat{b}$ ($\hbar \equiv 1$) describes the cavity
and mechanical oscillation with their frequency $\omega_c$ and $\omega_m$,
respectively, and
\begin{eqnarray}
\vspace{-0.3cm} H_S(t)&=&-\sqrt{2}g\{\cos(\omega_m t) \hat{x}%
_m+\sin(\omega_m t)\hat{p}_m\}\hat{a}^\dagger \hat{a}  \nonumber \\
&+&iE(t)(\hat{a}^\dagger e^{i\Delta_0 t}-\hat{a}e^{-i\Delta_0 t})  \label{HS}
\end{eqnarray}
inside the time-ordered exponential is the system Hamiltonian in the
interaction picture with respect to $H_0$, which is obtained by the transformation 
$H_S(t)=e^{iH_0t}H(t)e^{-iH_0t}$ on the Hamiltonian $H(t)=-g(\hat{b}+
\hat{b}^{\dagger})\hat{a}^{\dagger}\hat{a}+iE(t)(\hat{a}^{\dagger}e^{-i%
\omega_0t} -\hat{a}e^{i\omega_0t})$ of the OMS. 
In the above equation, $g$ is the optomechanical coupling constant, and $%
\Delta_0=\omega_c-\omega_0$ is the detuning of the drive's central frequency
from the cavity frequency. The dimensionless mechanical coordinate operator
and mechanical momentum operator are defined as $\hat{x}_{m}=(\hat{b}+\hat{b}%
^\dagger)/\sqrt{2}$ and $\hat{p}_{m}=-i(\hat{b}-\hat{b}^\dagger)/\sqrt{2}$,
respectively. The cavity (mechanical) damping at the rate $\kappa$ ($%
\gamma_m $) can be described in terms of a linear coupling between the
cavity (mechanical) mode with the stochastic Langevin noise operator $\hat{%
\xi}_c$ ($\hat{\xi}_m$) of the reservoir \cite{book}:
\begin{eqnarray}
H_D(t)=i\big(\sqrt{\kappa}\hat{a}^\dagger \hat{\xi}_c(t) +\sqrt{\gamma_m}%
\hat{b}^\dagger \hat{\xi}_m(t)\big )+H.c.  \label{HD}
\end{eqnarray}
The associated noises are assumed to be the white ones satisfying $\langle
\hat{\xi}_l(t) \hat{\xi}^{\dagger}_l(\tau) \rangle_R=(n_{l}+1)\delta(t-\tau)$
($l=c,m$) with the respective quanta number $n_{l}$ in thermal equilibrium.
Such approximation is valid for the mechanical reservoir given the quality
factor $\omega_m/\gamma_m\gg 1$ \cite{vitali}. Because the system-reservoir
coupling in (\ref{HD}) takes its form in the interaction picture with
respect to the total self oscillation Hamiltonian of both system and
reservoir, it should be added into the time-ordered exponential $\mathcal{T}
\exp\{-i\int_0^t d\tau H_S(\tau)\}$ in the interaction picture to construct
the evolution operator $U_S(t,0)=\mathcal{T}e^{-i\int_0^t d\tau \big(%
H_S(\tau)+H_D(\tau)\big)}$ for the combination of the OMS and its associated
reservoirs (its momentary action $U_S(t+dt,t)$ gives the exact Langevin
equation and master equation of the OMS) \cite{book}.

The development of the entanglement between the cavity and mechanical mode
is closely connected to the dynamical evolution of these modes. Their
evolution under $U_{S}(t,0)$ involves three non-commutative
processes---cavity drive, optomechanical coupling and dissipation, so it is
impossible to solve the system dynamics directly from this joint evolution
operator. Our method to reduce the intricacy is factorizing it into numerous
ones corresponding to relatively tractable processes \cite{bhe}. Here we
apply the technique to find a factorization that is suitable to study the
dynamically evolving Gaussian states. Our factorization is obtained as
\[
U_{S}(t,0)=U_{E}(t,0)U_{OM}(t,0)U_{K}(t,0)U_{D}(t,0),
\]%
where $U_{D}(t,0)=\mathcal {T}\exp \{-i\int_{0}^{t}d\tau H_{D}(\tau )\}$ (see
Appendix A for details). The effective Hamiltonian in the first operator $%
U_{E}(t,0)=\mathcal{T}\exp \{-i\int_{0}^{t}d\tau \tilde{H}_{E}(\tau )\}$ for
the pure cavity drive process takes the form
\[
\tilde{H}_{E}(\tau )=iE(\tau )e^{i\Delta _{0}\tau }\hat{A}^{\dagger }(t,\tau
)+H.c,
\]%
with $\hat{A}(t,\tau )=e^{-\frac{\kappa }{2}(t-\tau )}\hat{a}+\hat{n}%
_{c}(t,\tau )$ being the sum of the decayed cavity mode operator and the
colored cavity noise operator $\hat{n}_{c}(t,\tau )=\sqrt{\kappa }\int_{\tau
}^{t}d\tau ^{\prime }e^{-\kappa (\tau ^{\prime }-\tau )/2}\hat{\xi}_{c}(\tau
^{\prime })$. The third evolution operator is $U_{K}(t,0)=\mathcal{T}\exp
\{ig\int_{0}^{t}d\tau \hat{K}_{m}(t,\tau )\hat{A}^{\dagger }\hat{A}(t,\tau
)\}$, where $\hat{K}_{m}(t,\tau )=\cos (\omega _{m}\tau )\hat{X}_{m}(t,\tau
)+\sin (\omega _{m}\tau )\hat{P}_{m}(t,\tau )$ is a linear combination of
the mechanical operators $\hat{X}_{m}(t,\tau )=\hat{B}(t,\tau )+\hat{B}%
^{\dagger }(t,\tau )$ and $\hat{P}_{m}(t,\tau )=-i\hat{B}(t,\tau )+i\hat{B}%
^{\dagger }(t,\tau )$ from $\hat{B}(t,\tau )=e^{-\frac{\gamma _{m}}{2}%
(t-\tau )}\hat{b}+\hat{n}_{m}(t,\tau )$ and $\hat{n}_{m}(t,\tau )=\sqrt{%
\gamma _{m}}\int_{\tau }^{t}d\tau ^{\prime }e^{-\gamma _{m}(\tau ^{\prime
}-\tau )/2}\hat{\xi}_{m}(\tau ^{\prime })$. To the first order of the
optomechanical coupling constant $g$, the effective Hamiltonian in the
process $U_{OM}(t,0)=\mathcal{T}\exp \{-i\int_{0}^{t}d\tau \tilde{H}_{OM}(\tau
)\}$ of optomechanical coupling is
\begin{eqnarray}
\tilde{H}_{OM}(\tau ) &=&g\hat{K}_{m}(t,\tau )\big(\hat{A}^{\dagger }(t,\tau
)D(\tau )+\hat{A}(t,\tau )D^{\ast }(\tau )  \nonumber \\
&+&|D(\tau )|^{2}\big),  \label{LOM}
\end{eqnarray}%
where
\begin{eqnarray}
D(\tau ) &=&e^{-\frac{\kappa }{2}(t-\tau )}\int_{0}^{\tau }dt^{\prime
}E(t^{\prime })e^{i\Delta _{0}t^{\prime }}e^{-\frac{\kappa }{2}(t-t^{\prime
})}  \nonumber \\
&+&\int_{0}^{\tau }dt^{\prime }[\hat{n}_{c}(t,t^{\prime }),\hat{n}%
_{c}^{\dagger }(t,\tau )]E(t^{\prime })e^{i\Delta _{0}t^{\prime }}.
\end{eqnarray}
Under the effective Hamiltonian in (\ref{LOM}), the OMS evolves according to the following differential equations:
which evolves the cavity and mechanical mode in terms of the
following differential equations:
\begin{eqnarray}
&&-i\frac{d\hat{a}}{d\tau }=ge^{-(\kappa +\gamma _{m})(t-\tau )/2}D(\tau
)(e^{-i\omega _{m}\tau }\hat{b}+e^{i\omega _{m}\tau }\hat{b}^{\dagger })
\nonumber \\
&+&ge^{-\kappa (t-\tau )/2}D(\tau )\cos (\omega _{m}\tau )(\hat{n}%
_{m}(t,\tau )+\hat{n}_{m}^{\dagger }(t,\tau ))  \nonumber \\
&+&ge^{-\kappa (t-\tau )/2}D(\tau )\sin (\omega _{m}\tau )(i\hat{n}%
_{m}(t,\tau )-i\hat{n}_{m}^{\dagger }(t,\tau )),  \nonumber \\
&&-i\frac{d\hat{b}}{d\tau }=ge^{-(\kappa +\gamma _{m})(t-\tau )/2}e^{i\omega
_{m}\tau }(D^{\ast }(\tau )\hat{a}+D(\tau )\hat{a}^{\dagger })  \nonumber \\
&+&ge^{i\omega _{m}\tau -\frac{\gamma _{m}}{2}(t-\tau )}\big(\hat{n}%
_{c}(t,\tau )D^{\ast }(\tau )+\hat{n}_{c}^{\dagger }(t,\tau )D(\tau )\big)
\nonumber \\
&+&ge^{i\omega _{m}\tau -\frac{\gamma _{m}}{2}(t-\tau )}|D(\tau )|^{2}.
\label{om}
\end{eqnarray}%
The $\hat{a}$ ($\hat{b}$) terms on the right side of (\ref{om}%
) are due to the beam-splitter (BS) action in the quadratic Hamiltonian (\ref%
{LOM}), and the $\hat{a}^{\dagger }$ ($\hat{b}^{\dagger }$) terms reflect
the coexisting squeezing (SQ) action.

Next we start with an initial OMS state $\rho (0)$ in thermal equilibrium
with the environment, i.e. $\rho (0)$ is a Gaussian state as the product of
a cavity vacuum and a finite temperature mechanical thermal state. This
initial state becomes entangled under optomechanical coupling. Its evolution
can be studied by successively acting each factor in the factorized form $%
U_{E}(t,0)U_{OM}(t,0)U_{K}(t,0)U_{D}(t,0)$ of the joint evolution operator $%
U_{S}(t,0)$ on the total initial state
$\chi (0)=\rho (0)R(0)$,
in which $R(0)$ denotes the reservoir state in thermal equilibrium with $%
\rho (0)$. One has $U_{D}(t,0)\chi (0)U_{D}^{\dagger }(t,0)=\chi (0)$ since,
under thermal equilibrium, the system-reservoir coupling in (\ref{HD}) does
not change the state $\rho (0)$ (see Appendix B for details), and $%
U_{K}(t,0) $ also keeps $\chi (0)$ invariant because $\hat{A}(t,\tau
)|0\rangle _{C}=0$ for the combined initial vacuum state $|0\rangle _{C}$ of
the cavity and its zero temperature reservoir. Thus the expectation values
of system operators $\hat{O}(t)$ reduce to the following trace over system
and reservoir degrees of freedom (see Appendix B for details):
\begin{eqnarray}
\langle \hat{O}(t)\rangle &=&\mbox{Tr}_{S\otimes R}\big\{U_{OM}^{\dagger
}(t,0)U_{E}^{\dagger }(t,0)\hat{O}U_{E}(t,0)U_{OM}(t,0)  \nonumber \\
&\times &\chi (0)\big\}.\vspace{-0.3cm}  \label{main}
\end{eqnarray}%
In weak coupling regime where the Hamiltonian of $U_{OM}(t,0)$ takes the
form in (\ref{LOM}), the evolved OMS is preserved to be in Gaussian 
state, because the state
$\mbox{Tr}_R\{U_E(t,0)U_{OM}(t,0)\chi(0)U^%
\dagger_{OM}(t,0)U_E^\dagger(t,0)\}$ 
of the evolved OMS is only determined by
the quadratic Hamiltonian in $U_{OM}(t,0)$ and the displacement Hamiltonian in $U_{E}(t,0)$. 

\begin{figure}[b!]
\vspace{-0cm} \centering
\epsfig{file=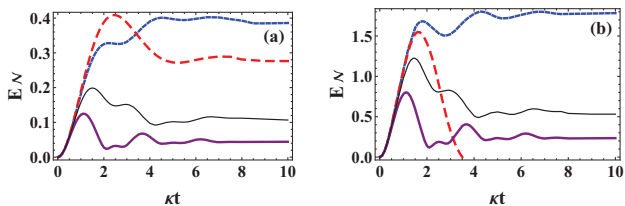,width=1.0\linewidth,clip=} {\vspace{-0.5cm}} \vspace{%
0cm}
\caption{(color online) Evolution of entanglement for blue detuned CW
drives. (a) is obtained with the drive intensity $E/\protect\kappa=3\times
10^5$ while (b) is for a drive of $E/\protect\kappa=2\times 10^6$. The long
dashed (red) curve is for $\Delta_0=-0.5\protect\omega_m$, the short dashed
(blue) curve for $\Delta_0=-\protect\omega_m$, the thin solid (black) curve
for $\Delta_0=-1.5\protect\omega_m$, and the thick solid (purple) curve for $%
\Delta_0=-2\protect\omega_m$. Here $g/\protect\kappa=10^{-6}$, $\protect%
\omega_m/\protect\kappa=2.5$, $\protect\omega_m/\protect\gamma_m=10^7$, and $%
T=0$. The entanglement measure by $E_{\mathcal{N}}$ for these blue detuned
drives tends to a stable value with time. The maximum entanglement is
reached at the SQ resonant point $\Delta_0=-\protect\omega_m$. For the
stronger drive, the entanglement at the smaller detuning $\Delta_0=-0.5%
\protect\omega_m$ dies a sudden death after a finite time period. }
\end{figure}

\section{OMS entanglement}

\subsection{Entanglement evolution under CW drive}

The entanglement of the evolved Gaussian states can be quantified by the logarithmic negativity $E_{\mathcal{N}%
} $ \cite{adesso}. One should consider the correlation matrix (CM) with the
elements
\[
\hat{V}_{ij}(t)=1/2\langle \hat{u}_{i}\hat{u}_{j}+\hat{u}_{j}\hat{u}%
_{i}\rangle -\langle \hat{u}_{i}\rangle \langle \hat{u}_{j}\rangle ,
\]%
where $\hat{\vec{u}}=(\hat{x}_{c}(t),\hat{p}_{c}(t),\hat{x}_{m}(t),\hat{p}%
_{m}(t))^{T}$, for the calculation of $E_{\mathcal{N}}$ (see Appendix C for
details). Each entry of the CM can be calculated following (\ref{main}) with
$\hat{O}=\hat{u}_{i}\hat{u}_{j}+\hat{u}_{j}\hat{u}_{i}$, etc. 

We first illustrate the real-time evolution of OMS entanglement under
the CW drives of different detuning. The first example we present in Fig. 2
is the entanglement evolution under blue detuned CW drives. The entanglement
values measured by $E_{\mathcal{N}}$ become stable with time and, at the SQ
resonant point $\Delta_0=-\omega_m$, the steady entanglement reaches the
maximum. Unlike the stationary entanglement between the fluctuations $\delta%
\hat{a}$ and $\delta\hat{b}$ under a SQ resonant drive, which is upper
bounded by $E_{\mathcal{N}}=\ln 2\approx 0.693$ due to the limitation of classical steady state conditions 
(see Eq. (\ref{condi}) below) \cite{vitali-08}, the evolved entanglement between the cavity mode $\hat{a}$ and mechanical
mode $\hat{b}$ themselves can be well beyond this limit (see Fig. 2(b)).
Compared with the blue detuned regime, the entanglement of the red detuned
regime shown in Fig. 3 is lower. This reflects the difference of the BS
action from the SQ action in creating the optomechanical entanglement.

\begin{figure}[t!]
\vspace{-0cm} \centering
\epsfig{file=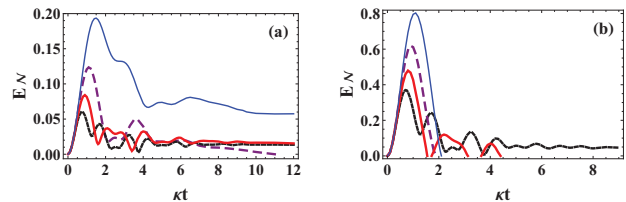,width=1.0\linewidth,clip=} {\vspace{-0.5cm}} \vspace{%
0cm}
\caption{(color online) Evolution of entanglement for red detuned CW drives.
(a) is obtained with the same system parameters as in Fig. 2(a), while (b)
is found with the same parameters as in Fig. 2(b). The thin solid (blue)
curve is for $\Delta_0=0.5\protect\omega_m$, and the long dashed (purple)
curve for $\Delta_0=\protect\omega_m$, the solid (red) curve for $%
\Delta_0=1.5\protect\omega_m$, and the short dashed (black) curve for $%
\Delta_0=2\protect\omega_m$. The entanglement dies earlier for a detuning
closer to the BS resonant point $\Delta_0=\protect\omega_m$ or the stronger
drive. Given the stronger drive in (b), the entanglement at $\Delta_0=1.5%
\protect\omega_m$ exhibits sudden death and revival. }
\end{figure}

\subsection{Quantum noise effect}

The exact degree of entanglement is determined by two competitive
factors---the direct BS and SQ action on the initial quantum state $\rho(0)$
of OMS, and the noise drives depending on the drive detuning and intensity.
Given a CW drive, the noise drive terms in (\ref{om}) are magnified by the
functions with the modulo $|D(\tau)|=E/\sqrt{0.25\kappa^2+\Delta_0^2}$,
indicating their more significant effect at a small detuning $\Delta_0$ or
with a stronger drive intensity $E$. In what follows, we illustrate the
noise effect as a function of time and of different system parameters.

\begin{figure}[b!]
\vspace{-0cm} \centering
\epsfig{file=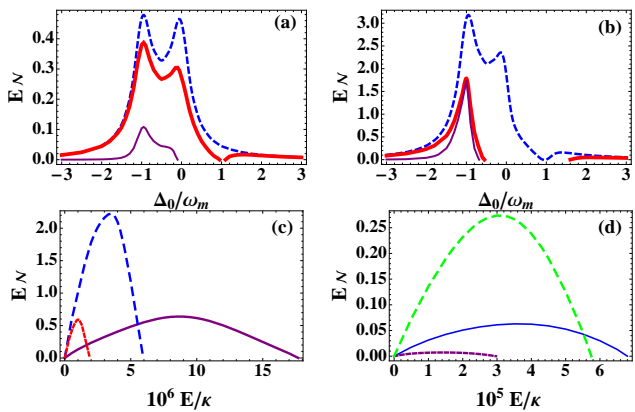,width=1.0\linewidth,clip=} {\vspace{-0.2cm}} \vspace{%
0cm}
\caption{(color online) (a)-(b): Entanglement versus detuning. (a) is under
the same conditions as in Figs. 2(a) and 3(a), and (b) corresponds to the
situation in Figs. 2(b) and 3(b). The long dashed (blue) curves show the
entanglement obtained at zero temperature without the noise drive terms in (%
\protect\ref{om}), while the thick solid (red) curves include the effect of
the noise drive terms at zero temperature. In (a) the zero temperature
entanglement is eliminated around $\Delta _{0}=\protect\omega _{m}$. The
thin solid (purple) curves in (a) and (b) give the exact degree of
entanglement at the temperature corresponding to $n_{m}=10^{4}$. (c)-(d):
Entanglement versus cavity drive intensity. In (c), the long-dashed curve
(blue) stands for $\Delta _{0}=-\protect\omega _{m}$, and the short-dashed
curve (red) for $\Delta _{0}=-0.5\protect\omega _{m}$, and the solid
(purple) curve for $\Delta _{0}=-2\protect\omega _{m}$. In (d), the
long-dashed curve (green) stand for $\Delta _{0}=0$, the short-dashed
(purple) curve for $\Delta _{0}=\protect\omega _{m}$, and the solid (blue)
curve for $\Delta _{0}=0.5\protect\omega _{m}$. Except for the thin solid
(purple) curves about finite temperature entanglement in (a) and (b), all
plots are found for the initial temperature $T=0$ at the moment $\protect%
\kappa t=15$, when the concerned entanglement has stabilized. }
\end{figure}

First, the entanglement for some values of detuning in Figs. 2 and 3 will
die at a finite time. The phenomenon that entanglement is killed by noise in
this way is known as entanglement sudden death (ESD) \cite{ESD,ESD2}. The
system evolution according to (\ref{om}) provides a model in which the ESD
for the continuous variable states is caused by the colored noises ($\hat{n}%
_c(t,\tau)$, $\hat{n}_m(t,\tau)$ and their conjugates on the right side of (%
\ref{om})) rather than the white noises in many other examples (see the
references in \cite{ESD2}). In this situation the noise effect can be so
significant that this type of ESD happens while the optomechanical coupling
exists all the time. Interestingly, the entanglement under some drives, e.g.
$\Delta_0=1.5\omega_m$ in Fig. 3(b), can also revive from time to time
during evolution.

\begin{figure}[t!]
\vspace{-0cm} \centering
\epsfig{file=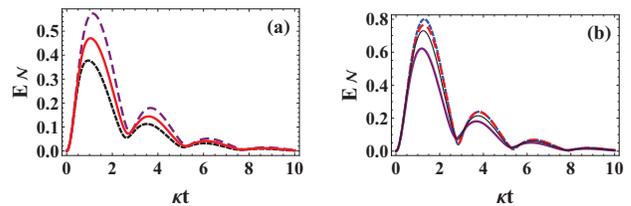,width=1.0\linewidth,clip=} {\vspace{-0.2cm}} \vspace{%
0cm}
\caption{(color online) Evolution of entanglement under the pulsed drive $%
E(t)=Ee^{-\Delta \protect\omega ^{2}t^{2}}$ with $\Delta \protect\omega =%
\protect\omega _{m}$. (a) Red detuned central frequencies. The long dashed
(purple) curve is for $\Delta _{0}=\protect\omega _{m}$ of the central
frequency, the solid (red) curve for $\Delta _{0}=1.5\protect\omega _{m}$,
and the short dashed (black) curve for $\Delta _{0}=2\protect\omega _{m}$.
(b) Blue detuned and resonant central frequencies. The long dashed (red)
curve is for $\Delta _{0}=0$, the short dashed (blue) curve for $\Delta
_{0}=-\protect\omega _{m}$, the thin solid (black) curve for $\Delta
_{0}=-1.5\protect\omega _{m}$, and the thick solid (purple) curve for $%
\Delta _{0}=-2\protect\omega _{m}$. The system parameters are $g/\protect%
\kappa =10^{-6}$, $E/\protect\kappa =2\times 10^{6}$, $\protect\omega _{m}/%
\protect\kappa =2.5$, $\protect\omega _{m}/\protect\gamma _{m}=10^{7}$, and $%
T=0$. The entanglement measured by $E_{\mathcal{N}}$ goes down at almost
same pace for the drives of different central frequency detuning. }
\end{figure}

Fig. 4 shows the magnitude of the noise correction to optomechanical
entanglement in the system parameter space. Given the same drive
intensities, the relations between the entanglement and drive detuning after
sufficiently long interaction time are shown in Figs.4(a)-4(b). With the
increase of cavity drive intensity, the entanglement in a more extended
detuning range around the BS resonant point $\Delta_0=\omega_m$ will be
eliminated by the quantum noises. The overall tendency of the entanglement
change with the drive intensity for various drive detuning values is
described in Figs. 4(c)-4(d). The plots in these figures show a competition
between the effective coupling $gD(t)$ and the noise drives [see the
respective terms in (\ref{om})] in affecting the degree of entanglement. The
entanglement reaches the maximum at a certain drive intensity $E$ determined
by the system parameters, instead of monotonically increasing with $E$ which
enhances the effective optomechanical coupling. Despite the existence of the
noises, the entanglement in the blue detuned regime can be high. The SQ
generated entanglement is also rather robust against temperature; see the
comparison in Figs. 4(a) and 4(b).

\subsection{Entanglement evolution under pulsed drive}

Finally, in Fig. 5, we provide an example of entanglement evolution for OMS driven by a
pulse. Pulsed optomechanics is a newly developed research field \cite%
{pulse2, Vanner1, Vanner2}. Here we use Gaussian pulses with the width $%
\omega _{m}$. Due to the contribution from a spectrum of frequencies, the
entanglement for the drives of different central frequency detuning evolves
similarly. Another noticeable feature is that, given the same system
parameters, the entanglement generated under the pulse could last even
longer than that of the CW ones. This can be explained by the contribution
from the frequency components outside the regime, in which the noise effect
quickly destroys the corresponding entanglement. Entanglement under pulsed
drive was also discussed in the approach based on classical mean motion
background \cite{pulse2}. Like in the CW cases, our results independent of
classical background are consequent upon the dynamics involving a
significantly different quantum noise effect.

\section{difference from entanglement of Fluctuations}

We are concerned with the regime of strong drive ($E/\kappa \gg 1$) and weak
optomechanical coupling ($g/\kappa \ll 1$) in the study of Gaussian state
entanglement for OMS. Starting from our initial OMS quantum state (the
cavity in a vacuum state and the mechanical oscillator in a thermal state),
such entanglement for the evolved quantum state develops as the
optomechanical coupling in Fig. 1 starts with the cavity field being built
up by an external drive $E(t)$. Meanwhile, the generated entanglement is
also weakened or even destroyed by the noise drives. The entanglement in the
same regime was well studied in the fluctuation expansion approach \cite%
{vitali, pater, vitali-07,vitali-08, h-p-08, galve, zou, abdi, G}. 
The steady entanglement of the cavity and mechanical fluctuation is based on the classical steady state
of OMS, the existence of which is determined by 
Routh-Hurwitz criterion \cite{RH} in terms of the following inequalities \cite{vitali}:
\begin{widetext}
\begin{eqnarray}
s_{1}&=&2\gamma _{m}\kappa \{[\kappa ^{2}+\left( \omega _{m}-\Delta \right)
^{2}][\kappa ^{2}+\left( \omega _{m}+\Delta \right) ^{2}]+\gamma
_{m}[(\gamma _{m}+2\kappa )   \left( \kappa ^{2}+\Delta ^{2}\right) +2\kappa \omega
_{m}^{2}]\}+\Delta \omega _{m}G^{2}(\gamma _{m}+2\kappa )^{2}>0,\nonumber\\
s_{2}&=&\omega _{m}\left( \kappa ^{2}+\Delta ^{2}\right) -G^{2}\Delta >0,
\label{condi}
\end{eqnarray}
\end{widetext}
with $G=\sqrt{2}g\alpha _{s}$ and $\Delta =\Delta _{0}-g^{2}|\alpha
_{s}|^{2}/\omega _{m}$ expressed in terms of the cavity field amplitude $\alpha
_{s}=E/(\kappa +i\Delta )$ as the stationary solution to the Langevin
equation. The workable regime for the fluctuation expansion approach is depicted with these conditions in Fig. 6. 
\begin{figure}[t]
\vspace{-0cm} \centering
\epsfig{file=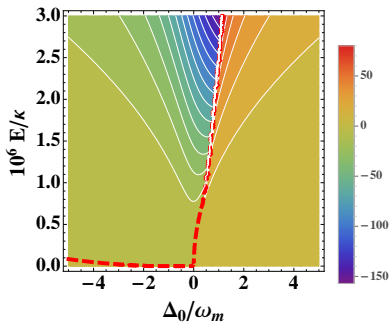,width=0.6\linewidth,clip=} {\vspace{-0.2cm}} \vspace{%
0cm}
\caption{(color online) The workable regime for the standard fluctuation expansion approach. The scale represents the value of $s_1$ 
defined in (\ref{condi}). The second condition $s_2>0$ is always satisfied in the concerned regime. The fluctuation expansion approach works in the regime where the scale takes the positive value, which is separated from the other part of the whole parameter space by the boundary of the dashed line. On the other hand, 
the approach presented in this paper is valid at any point of the parameter space. }
\end{figure}
Both of the approaches work in the red detuned regime. Fig. 7, however, shows that even in this common regime the entanglement 
between the fluctuations can be very different from the OMS entanglement discussed in this paper.
\begin{figure}[h!]
\vspace{-0cm} \centering
\epsfig{file=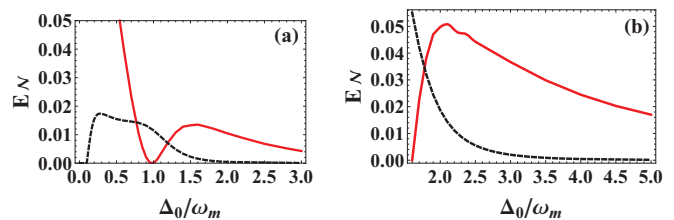,width=1.05\linewidth,clip=} {\vspace{-0.2cm}} \vspace{%
0cm}
\caption{Comparison of entanglement obtained in our approach (solid curve)
and in the fluctuation expansion approach of \protect\cite{vitali} (dashed
curve). (a) is obtained with the drive intensity $E/\protect\kappa =3\times
10^{5}$ while (b) is for a drive of $E/\protect\kappa =2\times 10^{6}$. The
system parameters are $g/\protect\kappa =10^{-6}$, $\protect\omega _{m}/%
\protect\kappa =2.5$, $\protect\omega _{m}/\protect\gamma _{m}=10^{7}$, and $%
T=0$. The entanglement represented by the solid curves is obtained at $\protect\kappa t=15$. }
\label{Fig:u}
\end{figure}

\begin{figure}[t]
\vspace{-0cm} \centering
\epsfig{file=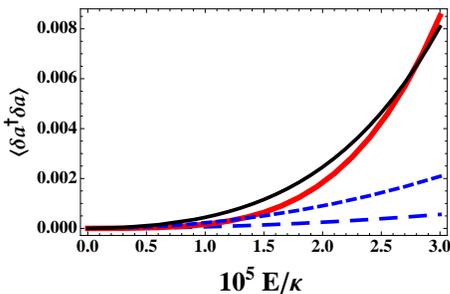,width=0.69\linewidth,clip=} {\vspace{-0.2cm}} \vspace{%
-0.3cm}
\caption{(color online) Comparison of the cavity fluctuation $\langle
\protect\delta \hat{a}^{\dagger }\protect\delta \hat{a}\rangle =\langle (%
\hat{a}^{\dagger }-\langle \hat{a}^{\dagger }\rangle )(\hat{a}-\langle \hat{a%
}\rangle )\rangle =\langle \hat{a}^{\dagger }\hat{a}\rangle -\langle \hat{a}%
^{\dagger }\rangle \langle \hat{a}\rangle $ at zero temperature obtained in
the approach of the present work (solid curves) and in the fluctuation
expansion approach of Ref. \protect\cite{vitali} (dashed curves). The plots
for our approach are obtained at the time $\protect\kappa t=15$. The thick
solid (red) and long dashed curves compare the fluctuation at the detuning $%
\Delta _{0}=\protect\omega _{m}$, while the thin solid (black) and short
dashed compare that at $\Delta _{0}=0.5\protect\omega _{m}$.}
\end{figure}

As we mentioned at the beginning, the fluctuation expansion approach works
with approximating the OMS operators with the sum of their mean values
following classical dynamics without noise drives and the fluctuations
evolving according to quantum mechanics. Then the system operators in the
system-reservoir coupling of (\ref{HD}) are replaced by their fluctuations,
so that only delta-function correlated Langevin noises $\hat{\xi}_{c}$ and $%
\hat{\xi}_{m}$ independent of cavity drive detuning and intensity are
relevant to the linearized dynamics about the fluctuations and their
entanglement. Instead, in our fully quantum approach, the linearized
dynamics for the system operators in weak coupling regime involves the
magnified noise drives due to the cubic term $-g(\hat{b}+\hat{b}^{\dagger })%
\hat{a}^{\dagger }\hat{a}$ of the original OMS Hamiltonian; see Eq. (\ref{om}%
). The difference of the quantum noise effects is expected to be
experimentally tested by the measurement of cavity fluctuation amplitude. As
illustrated in Fig. 8, the cavity fluctuations found in the different
approaches drastically deviate with drive intensity. This phenomenon also
indicates the distinct quantum states due to the different linearized
dynamics. Our concerned OMS quantum states and those of the fluctuations
around classical steady states can be seen to be different from their CMs,
which are in one-to-one correspondence to the respective Gaussian states.
The entanglement for the evolved states of fully quantum OMS can thus
significantly differ from that previously considered in the fluctuation
expansion approach.

\section{Conclusion}

In conclusion, we have studied the dynamically generated entanglement of
quantum OMSs that are initially in thermal equilibrium with their
environment. The meaning of our research manifests in two aspects. First,
one sees that high and robust entanglement for fully quantum OMSs can be
generated with blue detuned drive. In contrast, the previous fluctuation
expansion approximation working under the classical steady state condition
specifies an upper bound for the entanglement in blue detuned regime and
only focuses on the steady entanglement under red detuned drives. This
finding involving the different implementations is important to the related
experimental studies on OMSs entering quantum regime. Second, our fully
quantum dynamical approach shows that the noise effect on a quantum OMS
drastically differs from that affecting the cavity and mechanical
fluctuation considered in the previous approach, though the system dynamics
is linearized for weak optomechanical coupling in both approaches. In the
regime where the magnified noise effect is significant, complicated
evolution patterns such as entanglement sudden death and revival exist for
our concerned macroscopic entanglement. Such non-trivial quantum noise
effect can also exist in other quantum nonlinear systems.

\vspace{2cm}
\begin{acknowledgments}
B.H. thanks M. Hillery for helpful conversations. This work was supported by
AITF and NSERC. Q. L. acknowledges the support by NSFC(No. 11005040),
NCETFJ(No. 2012FJ-NCET-ZR04), PPYMTSTRHU (No. ZQN-PY113) and CSC.
\end{acknowledgments}

\vspace{1.5cm}

\appendix
\begin{widetext}
\section{Factorization of Joint System-Reservoir Evolution Operator}
\renewcommand{\theequation}{A-\arabic{equation}}
 \setcounter{equation}{0}

Our discussion is based on the two following factorizations for a unitary evolution operator
$U(t,0)=\mathcal{T}\exp \{-i\int_0^t d\tau \big(H_1(\tau)  +H_2 (\tau)\big)\}$ involving two processes
described by $H_1(t)$ and $H_2(t)$, respectively:
 \begin{eqnarray}
\mathcal{T}e^{-i\int_0^t d\tau (H_1(\tau) +H_2 (\tau))}
= \mathcal{T}e^{-i\int_0^t d\tau H_1(\tau)  }~\mathcal{T}e^{-i\int_0^t d\tau V^{\dagger}_1(\tau,0)H_2(\tau)V_1(\tau,0)},
\label{a}
\end{eqnarray}
and
\begin{eqnarray}
\mathcal{T}e^{-i\int_0^t d\tau (H_1(\tau)  +H_2 (\tau))}
=\mathcal{T}e^{-i\int_0^t d\tau V_2(t,\tau )H_1(\tau)V^{\dagger}_2(t,\tau) }~\mathcal{T}e^{-i\int_0^t d\tau H_2(\tau)},
\label{b}
\end{eqnarray}
where $V_k(t,\tau)=\mathcal{T}\exp \{-i\int_\tau^t d\tau' H_k(\tau')\}$ for $k=1,2$.
The operator $U(t,0)$ is the solution to the differential equations $d U/dt=-i\big(H_1(t)+H_2 (t)\big)\hat{U}(t)$,
while $V_1(t,0)=\mathcal{T}\exp \{-i\int_0^t d\tau H_1(\tau) \}$ is the solution to
$d V_1/dt=-i H_1(t)V_1(t)$. The initial condition for the differential equations is $U(0,0)=V_1(0,0)=I$, the identity operator.
The differential of $W(t,0)=V_1^{\dagger}(t,0)U(t,0)$ with respect to $t$ gives
\begin{eqnarray}
\frac{d W}{dt}&=&-V_1^{\dagger}\frac{d V_1}{dt}V_1^{\dagger}U+V_1^{\dagger}\frac{d U}{dt}=iV_1^{\dagger}H_1V_1\hat{V}_1^{\dagger}\hat{U}-iV_1^{\dagger}(H_1+H_2)\hat{U}=-iV_1^{\dagger}H_2 V_1 W.
\end{eqnarray}
One has the solution to the above differential equation as $W(t,0)=\mathcal{T}\exp \{-i\int_0^t d\tau V_1^{\dagger}(\tau,0)H_2(\tau)V_1(\tau,0)\}$, thus proving the factorization in (\ref{a}).
By exchanging $H_1(t)$ and $H_2(t)$ in (\ref{a}), one has the factorization of the operator $U(t,0)$ as
$$V_2(t,0)~\mathcal{T}e^{-i\int_0^t d\tau V^{\dagger}_2(\tau,0)H_1(\tau)V_2(\tau,0)}=V_2(t,0)~\mathcal{T}e^{-i\int_0^t d\tau V^{\dagger}_2(\tau,0)H_1(\tau)V_2(\tau,0)}V^\dagger_2(t,0)V_2(t,0).$$
Because $V_2(t,0)$ is a unitary operation, one can rewrite the right side of the above as $\mathcal{T}e^{-i\int_0^t d\tau V_2(t,\tau)H_1(\tau)V^\dagger_2(t,\tau)}V_2(t,0)$, giving the form
in Eq.(\ref{b}). Here we have used the relation $V_2(t,0)V^{\dagger}_2(\tau,0)=V_2(t,\tau)$.

We first apply Eq. (\ref{b}) to factorize $U_D(t,0)=\mathcal {T}\exp\{-i\int_{0}^t d\tau H_D(\tau)\}$ out of the system-reservoir evolution operator $U_S(t,0)=\mathcal{T}\exp\{-i\int_0^t d\tau \big(H_S(\tau)+H_D(\tau)\big)\}$, where $H_S(\tau)$ and $H_D(\tau)$ are given in Eqs. (1) and (2) of the main text, respectively.
In this way one has
\begin{eqnarray}
U_S(t,0)=\mathcal{T}\exp\{-i\int_0^t d\tau U_D(t,\tau)H_S(\tau)U_D^\dagger(t,\tau)\}
~\mathcal {T}\exp\{-i\int_{0}^t d\tau H_D(\tau)\}.
\label{one}
\end{eqnarray}
The cavity mode operator $\hat{a}$ in $H_S(\tau)$ is transformed to
\begin{eqnarray}
U_D(t,\tau)\hat{a}U_D^\dagger(t,\tau)=e^{-\frac{\kappa}{2}(t-\tau)}\hat{a}+\hat{n}_c(t,\tau)\equiv \hat{A}(t,\tau)
\end{eqnarray}
in $U_D(t,\tau)H_S(\tau)U_D^\dagger(t,\tau)$, and the mechanical mode operator is transformed to
\begin{eqnarray}
U_D(t,\tau)\hat{b}U_D^\dagger(t,\tau)=e^{-\frac{\gamma_m}{2}(t-\tau)}\hat{b}+\hat{n}_m(t,\tau)\equiv\hat{B}(t,\tau),
\end{eqnarray}
where $\hat{n}_c(t,\tau)=\sqrt{\kappa}\int_{\tau}^t d\tau'e^{-\kappa(\tau'-\tau)/2}\hat{\xi}_c(\tau')$ and $\hat{n}_m(t,\tau)=\sqrt{\gamma_m}\int_{\tau}^t d\tau'e^{-\kappa(\tau'-\tau)/2}\hat{\xi}_m(\tau')$ \cite{bhe}.
The transformed operators satisfy the equal-time commutation relation $[\hat{A}(t,\tau), \hat{A}^\dagger(t,\tau)]=[\hat{B}(t,\tau),\hat{B}^\dagger(t,\tau)]=1$. Then the Hamiltonian in the first time-ordered exponential of (\ref{one}) becomes
\begin{eqnarray}
U_D(t,\tau)H_S(\tau)U_D^\dagger(t,\tau)=\big(iE(t)\hat{A}^\dagger(t,\tau) e^{i\Delta_0 t}-iE^\ast(t)\hat{A}(t,\tau)e^{-i\Delta_0 t})-g\hat{K}_m(t,\tau)\hat{A}^\dagger\hat{A}(t,\tau),
\label{ts}
\end{eqnarray}
where
$$\hat{K}_m(t,\tau)=\cos(\omega_m \tau) \big(\hat{B}_m(t,\tau)+\hat{B}^\dagger_m(t,\tau)\big)+\sin(\omega_m \tau)\big(-i\hat{B}_m(t,\tau)+i\hat{B}^\dagger_m(t,\tau)\big).$$

By using (\ref{a}) we factorize out the drive Hamiltonian in (\ref{ts}) as follows:
\begin{eqnarray}
&&\mathcal{T}\exp\{-i\int_0^t d\tau U_D(t,\tau)H_S(\tau)U_D^\dagger(t,\tau)\}\nonumber\\
&=& \mathcal{T}\exp\big\{-i\int_0^t d\tau\big(iE(t)\hat{A}^\dagger(t,\tau) e^{i\Delta_0 t}-iE^\ast(t)\hat{A}(t,\tau)e^{-i\Delta_0 t})~\mathcal{T}\exp\big\{ig\int_0^t d\tau U_E^\dagger(\tau,0)\hat{K}_m(t,\tau)\hat{A}^\dagger\hat{A}(t,\tau)U_E(\tau,0)\big\}\nonumber\\
&=&\mathcal{T}\exp\big\{-i\int_0^t d\tau\big(iE(t)\hat{A}^\dagger(t,\tau) e^{i\Delta_0 t}-iE^\ast(t)\hat{A}(t,\tau)e^{-i\Delta_0 t}\big)\big\}\nonumber\\
&\times & \mathcal{T}\exp\big\{ig\int_0^t d\tau \hat{K}_m(t,\tau)\big(\hat{A}^\dagger(t,\tau)+D^\ast(\tau)\big)\big(\hat{A}(t,\tau)+D(\tau)\big)\big\},
\label{two}
\end{eqnarray}
where $U_{E}(\tau,0)=\mathcal{T}\exp\{\int_0^\tau dt' E(t')e^{i\Delta_0 t'}\hat{A}^\dagger(t,t')-H.c.\}$. In (\ref{two}) the effect of $U_{E}(\tau,0)$ on the cavity operator $\hat{A}(t,\tau)$ is the displacement
\begin{eqnarray}
U_E^\dagger(\tau,0)\hat{A}(t,\tau)U_E(\tau,0)&=&\hat{A}(t,\tau)+e^{-\frac{\kappa}{2}(t-\tau)}\int_0^\tau dt' E(t')e^{i\Delta_0 t'}e^{-\frac{\kappa}{2}(t-t')}
+\int_0^\tau dt' \Gamma_c(t',\tau)E(t')e^{i\Delta_0 t'}\nonumber\\
&\equiv & \hat{A}(t,\tau)+D(\tau),
\label{displace}
\end{eqnarray}
where
$$\Gamma_c(t',\tau)=[\hat{n}_c(t,t'),\hat{n}_c^{\dagger}(t,\tau)]=e^{-\kappa(\tau-t')/2}-e^{-\kappa(t-\tau)/2}
e^{-\kappa(t-t')/2}.$$

The next step is to factorize the second time-ordered exponential in (\ref{two}) as follows:
\begin{eqnarray}
&&\mathcal{T}\exp\big\{ig\int_0^t d\tau \hat{K}_m(t,\tau)\big(\hat{A}^\dagger(t,\tau)+D^\ast(\tau)\big)\big(\hat{A}(t,\tau)+D(\tau)\big)\big\}\nonumber\\
&=&\mathcal{T}\exp\big\{ig U_K(t,\tau)\big(\hat{K}_m(t,\tau) \big(\hat{A}^{\dagger}(t,\tau)D(\tau)+\hat{A}(t,\tau)D^{\ast}
(\tau )+|D(\tau)|^2\big)U_K^\dagger(t,\tau)\big\}\nonumber\\
&\times &\mathcal{T}\exp\{ig\int_0^t d\tau \hat{K}_m(t,\tau)\hat{A}^{\dagger}\hat{A}(t,\tau)\},
\label{three}
\end{eqnarray}
where $U_K(t,\tau)=\mathcal{T}\exp\{ig\int_\tau^t dt' \hat{K}_m(t,t')\hat{A}^{\dagger}\hat{A}(t,t')\}$.
To the first order of $g$, the operation $U_K(t,\tau)$ in the first time-ordered exponential of the above equation transforms the mechanical operator as
\begin{eqnarray}
U_K(t,\tau)\hat{K}_m(t,\tau)U_K^\dagger(t,\tau)&=&\hat{K}_m(t,\tau)-2g \int_\tau^t d\tau' e^{-\gamma_m(\tau'-\tau)/2} \sin \omega_m(\tau-\tau')\hat{A}^{\dagger}(t,\tau')\hat{A}(t,\tau')+\cdots\nonumber\\
\end{eqnarray}
and the cavity operator $\hat{A}(t,\tau)$ as
\begin{eqnarray}
&&U_K(t,\tau)\hat{A}(t,\tau)U_K^\dagger (t,\tau)
=\hat{A}(t,\tau)-ig\int_\tau^t dt' e^{-\kappa(t'-\tau)/2}\hat{K}_m(t,t')\hat{A}(t,t')+\cdots
\end{eqnarray}
The neglected higher order terms in the above expansions are successively lowered by a small factor in the order of $g(t-\tau)$, because all terms in the expansions only contain the operators $\hat{K}_m$ and $\hat{A}$ not magnified by the drive intensity $E(t)$, so the dominant first order contribution leads to the effective optomechanical coupling Hamiltonian
\begin{eqnarray}
\tilde{H}_{OM}(\tau)=-g \hat{K}_m(t,\tau)
\big(D^\ast(\tau)\hat{A}(t,\tau)+D(\tau) \hat{A}^\dagger(t,\tau)+| D(\tau)|^2\big)
\end{eqnarray}
for the first time-ordered exponential in
(\ref{three}), which is defined as $U_{OM}(t,0)=\mathcal{T}\exp\{-i\int_0^t d\tau \tilde{H}_{OM}(\tau)\}$.
Now we have exactly factorized the joint evolution operator as
\begin{eqnarray}
U_S(t,0)=U_E(t,0)U_{OM}(t,0)U_K(t,0)U_D(t,0).
\label{fac}
\end{eqnarray}

\section{Expectation Value of System Operators}
\renewcommand{\theequation}{B-\arabic{equation}}
 \setcounter{equation}{0}

We apply the factorization of the joint evolution operator in (\ref{fac}) to find the expectation value of a system operator $\hat{O}$:
 \begin{eqnarray}
\mbox{Tr}_S\{\hat{O}\rho(t)\}&=&\mbox{Tr}_S\big\{\hat{O}~\mbox{Tr}_R\{U_E(t,0)U_{OM}(t,0)U_K(t,0)U_D(t,0)\rho(0)R(0)U^\dagger_D(t,0)U^\dagger_K(t,0)U^\dagger_{OM}(t,0)U^\dagger_E(t,0)\}\big\}\nonumber\\
&=&\mbox{Tr}_{S\otimes R}\big\{U^\dagger_{OM}(t,0)U^\dagger_{E}(t,0)\hat{O}U_{E}(t,0)U_{OM}(t,0)\big(U_K(t,0)U_D(t,0)\rho(0)R(0)U^\dagger_D(t,0)U^\dagger_K(t,0)\big)\big\}.
\label{exp}
\end{eqnarray}
The action $U_K(t,0)U_D(t,0)\rho(0)R(0)U^\dagger_D(t,0)U^\dagger_K(t,0)$ is on the
the product of the initial system state
\begin{eqnarray}
\rho(0)=|0\rangle_c\langle 0|\otimes \sum_{n=0}^\infty \frac{n_{m}^n}{(1+n_{m})^{n+1}}|n\rangle_m\langle n|\equiv |0\rangle_c\langle 0|\otimes \rho_m
\label{input}
\end{eqnarray}
and the associate reservoir state $R(0)$ in thermal equilibrium with $\rho(0)$, where $n_m$ is the thermal phonon number at the temperature $T$.

We first look at $U_D(t,0)\chi(0)U^\dagger_D(t,0)$, where $\chi(0)=\rho(0)R(0)$ and
$$U_D(t,0)=\mathcal{T}\exp\big\{\int_0^td\tau\big(\sqrt{\gamma_m}\hat{b}^\dagger \hat{\xi}_m(\tau)-\sqrt{\gamma_m}\hat{b} \hat{\xi}^\dagger_m(\tau)\big)\big\}
~\mathcal{T}\exp\big\{\int_0^td\tau\big(\sqrt{\kappa}\hat{a}^\dagger \hat{\xi}_c(\tau)-\sqrt{\kappa}\hat{a} \hat{\xi}^\dagger_c(\tau)\big)\big\}.$$
The second operator of the cavity and vacuum reservoir coupling does not change $\chi(0)$ because
\begin{eqnarray}
\big(\sqrt{\kappa}\hat{a}^\dagger \hat{\xi}_c(\tau)-\sqrt{\kappa}\hat{a} \hat{\xi}^\dagger_c(\tau)\big)|0\rangle_C=0
\label{vc}
\end{eqnarray}
for the product state $|0\rangle_C$ of the cavity vacuum and its associate vacuum reservoir. If the action of the first operator involving mechanical mode and mechanical reservoir coupling changes the joint initial state $\chi(0)$, the system state
$$\tilde{\rho}(t)=\mbox{Tr}_R \big\{\mathcal{T}e^{\int_0^td\tau\{\sqrt{\gamma_m}\hat{b}^\dagger \hat{\xi}_m(\tau)-\sqrt{\gamma_m}\hat{b} \hat{\xi}^\dagger_m(\tau)\}} \chi(0)\mathcal{T}e^{-\int_0^td\tau \{\sqrt{\gamma_m}\hat{b}^\dagger \hat{\xi}_m(\tau)-\sqrt{\gamma_m}\hat{b} \hat{\xi}^\dagger_m(\tau)\}}\big\}$$
evolved under such coupling will be different from $\rho(0)$. The system quantum state $\tilde{\rho}(t)$ is the solution to the master equation
\begin{eqnarray}
\dot{\tilde{\rho}}&=&\gamma_m(n_{th}+1)\big\{\hat{b}\tilde{\rho}(t)\hat{b}^{\dagger}-\frac{1}{2}\tilde{\rho}(t)\hat{b}^{\dagger}\hat{b}-\frac{1}{2}\hat{b}^{\dagger}\hat{b}
\tilde{\rho}(t)\big\}
+\gamma_m n_{th}\big\{\hat{b}^\dagger\tilde{\rho}(t)\hat{b}-\frac{1}{2}\tilde{\rho}(t)\hat{b}\hat{b}^\dagger-\frac{1}{2}\hat{b}\hat{b}^\dagger
\tilde{\rho}(t) \big\}
\label{master}
\end{eqnarray}
in Lindblad form \cite{book}. The initial state for the above master equation is $\tilde{\rho}(0)=\rho_m$, and $n_{th}$ is the thermal quantum number of the reservoir. Here we assume the possible non-equilibrium between system and reservoir, so that $n_{th}$ could be different from $n_m$ (in the main text we only consider the situation of thermal equilibrium). This master equation can be exactly solved by the super-operator technique \cite{a-m} as
\begin{eqnarray}
\tilde{\rho}(t)=\sum_{n=0}^\infty\frac{\big(n_{th}+(n_m-n_{th})e^{-\gamma_m t/2}\big)^n}{\big(1+n_{th}+(n_m-n_{th})e^{-\gamma_m t/2}\big)^{n+1}}|n\rangle_m\langle n|.
\end{eqnarray}
If the system and reservoir is in thermal equilibrium, i.e. $n_{th}=n_m$, the above state will be
$\rho_m$ constantly with time.
Under this condition, therefore, the operation $U_{D}(t,0)$ keeps the joint initial state $\chi(0)$ invariant.
Moreover, similar to (\ref{vc}), one has $U_K(t,0)\chi(0)U_K^\dagger(t,0)=\chi(0)$.
Thus the system operator expectation value in (\ref{exp}) will reduce to the form in (7) of the main text.

\section{Calculation of Entanglement Measured by Logarithmic Negativity}
\renewcommand{\theequation}{C-\arabic{equation}}
\renewcommand{\thefigure}{C-\arabic{figure}}
\setcounter{equation}{0}
\setcounter{figure}{0}

The entanglement of bipartite Gaussian states is quantified via the correlation matrix
\begin{eqnarray}
\hat{V}= \left(
\begin{array}
[c]{cc}%
 \hat{A} & \hat{C} \\
 \hat{C}^T &  \hat{B}
\end{array}
\right).
\label{corr-matrix}
\end{eqnarray}
with the elements $\hat{V}_{ij}(t)=0.5\langle \hat{u}_i\hat{u}_j+\hat{u}_j\hat{u}_i\rangle-\langle \hat{u}_i\rangle\langle \hat{u}_j\rangle$, where $\hat{\vec{u}}=(\hat{x}_c(t),\hat{p}_c(t),\hat{x}_m(t),\hat{p}_m(t))^T$. The logarithmic negativity as a measure for the entanglement is given as \cite{v-w,adesso}
\begin{eqnarray}
E_{\cal N}=\mbox{max}[0, -\ln 2\eta^{-}],
\end{eqnarray}
where
\begin{eqnarray}
\eta^{-}=\frac{1}{\sqrt{2}}\sqrt{\Sigma-\sqrt{\Sigma^2-\mbox{det}\hat{V}}}
\end{eqnarray}
and
\begin{eqnarray}
\Sigma=\mbox{det}\hat{A}+\mbox{det}\hat{B}-2\mbox{det}\hat{C}.
\end{eqnarray}

$U_E(t,0)$ in (\ref{exp}) does not contribute to the correlation matrix elements. Given the quadratic Hamiltonian $H_{OM}$ in (6) of the main text, the operation $U_{OM}$ transforms the
vector $(\hat{x}_c(t),\hat{p}_c(t),\hat{x}_m(t),\hat{p}_m(t))^T$ in terms of the following linear differential equation:
\begin{eqnarray}
\frac{d}{d\tau}\left(
\begin{array}
[c]{c}%
\hat{x}_c\\
\hat{p}_c\\
\hat{x}_m\\
\hat{p}_m
\end{array}
\right)&=&\left(
\begin{array}
[c]{cccc}%
 0 & 0 & l_3(t,\tau) & l_4 (t,\tau) \\
 0 &  0 & l_1(t,\tau) & l_2(t,\tau) \\
-l_2 (t,\tau) & l_4(t,\tau) & 0 &  0\\
l_1(t,\tau) &-l_3(t,\tau) & 0 & 0
\end{array}
\right)\left(
\begin{array}
[c]{c}%
\hat{x}_c \\
\hat{p}_c\\
\hat{x}_m \\
\hat{p}_m
\end{array}
\right)+\left(
\begin{array}
[c]{c}%
\hat{f}_1 \\
\hat{f}_2\\
\hat{f}_3 \\
\hat{f}_4
\end{array}
\right)\nonumber\\
&\equiv & \frac{d}{d\tau}\hat{\vec{v}}=\hat{M}(t,\tau)\hat{\vec{v}}+\hat{\vec{f}}(t,\tau),
\label{VCM}
\end{eqnarray}
where
\begin{eqnarray}
l_1(t,\tau)&=&g e^{-\kappa (t-\tau)/2-\gamma_m(t-\tau)/2} \big(D(\tau)+D^\ast(\tau)\big)\cos(\omega_m\tau),\nonumber\\
l_2(t,\tau)&=&g e^{-\kappa (t-\tau)/2-\gamma_m(t-\tau)/2} \big(D(\tau)+D^\ast(\tau)\big)\sin(\omega_m\tau),\nonumber\\
l_3(t,\tau)&=&ig e^{-\kappa (t-\tau)/2-\gamma_m(t-\tau)/2}\big(D(\tau)-D^\ast(\tau)\big)
\cos(\omega_m\tau),\nonumber\\
l_4(t,\tau)&=&ig e^{-\kappa (t-\tau)/2-\gamma_m(t-\tau)/2} \big(D(\tau)-D^\ast(\tau)\big)
\sin(\omega_m\tau),
\end{eqnarray}
and
\begin{eqnarray}
\hat{f}_1(t,\tau)&=&\frac{i}{\sqrt{2}}g~ e^{-\kappa (t-\tau)/2}\big(D(\tau)-D^\ast(\tau)\big)\big\{\cos(\omega_m\tau)\big(\hat{n}_m(t,\tau)+\hat{n}^\dagger_m(t,\tau)\big)-\sin(\omega_m\tau)\big(i\hat{n}_m(t,\tau)-i\hat{n}^\dagger_m(t,\tau)\big)\big\},\nonumber\\
\hat{f}_2(t,\tau)&=& \frac{1}{\sqrt{2}}ge^{-\kappa (t-\tau)/2}\big(D(\tau)+D^\ast(\tau)\big)\big\{\cos(\omega_m\tau)\big(\hat{n}_m(t,\tau)+\hat{n}^\dagger_m(t,\tau)\big)-\sin(\omega_m\tau)\big(i\hat{n}_m(t,\tau)-i\hat{n}^\dagger_m(t,\tau)\big)\big\},\nonumber\\
\hat{f}_3(t,\tau)&=&-g\big(\hat{n}_c(t,\tau)D^{\ast}(\tau)+\hat{n}^\dagger_c(t,\tau)D(\tau)+|D(\tau)|^2\big)
e^{-\gamma_m(t-\tau)/2}\sin(\omega_m\tau),\nonumber\\
\hat{f}_4(t,\tau)&=&~g\big(\hat{n}_c(t,\tau)D^{\ast}(\tau)+\hat{n}^\dagger_c(t,\tau)D(\tau)+|D(\tau)|^2\big)e^{-\gamma_m(t-\tau)/2}\cos(\omega_m\tau).
\label{noise}
\end{eqnarray}
In the above the terms containing $\hat{n}_c$, $\hat{n}_m$ and their conjugates contribute to the correlation matrix (\ref{corr-matrix}), and the pure drive terms proportional to $|D(\tau)|^2$ do not contribute to
$\hat{V}$, but they affect the system mean motion $\langle \hat{v}_i(t)\rangle$.
The solution to (\ref{VCM}) is
 \begin{eqnarray}
\hat{\vec{v}}(t)=\mathcal{T}e^{\int_0^t d\tau \hat{M}(t,\tau)}\hat{\vec{v}}(0)+\mathcal{T}e^{\int_0^t d\tau \hat{M}(t,\tau)}\int_0^t
d\tau
(\mathcal{T}e^{\int_0^\tau d\tau' \hat{M}(t,\tau')})^{-1}\hat{\vec{f}}(t,\tau).
\label{sol}
\end{eqnarray}

\begin{figure}[t!]
\vspace{-0cm}
\centering
\epsfig{file=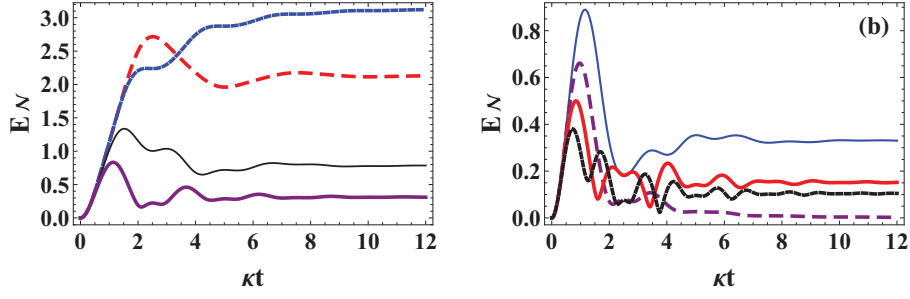,width=0.7\linewidth,clip=}
{\vspace{-0.5cm}\caption{\label{Fig:u} Entanglement evolution without quantum noise effect. (a) Blue detuned regime. The long dashed (red) curve is for $\Delta_0=-0.5\omega_m$, the short dashed (blue) curve for $\Delta_0=-\omega_m$, the thin solid (black) curve for $\Delta_0=-1.5\omega_m$, and the thick solid (purple) curve for $\Delta_0=-2\omega_m$. (b) Red detuned regime. The thin solid (blue) curve is for $\Delta_0=0.5\omega_m$, and the long dashed (purple) curve for $\Delta_0=\omega_m$, the solid (red) curve for $\Delta_0=1.5\omega_m$, and the short dashed (black) curve for $\Delta_0=2\omega_m$. The system parameters are $g/\kappa=10^{-6}$,
$E/\kappa=2\times 10^6$, $\omega_m/\kappa=2.5$, $\omega_m/\gamma_m=10^7$, and $T=0$. The plots in (a) and
(b) correspond to those in Fig. 1(b) and Fig. 2(b) of the main text, respectively. Here the entanglement without the effect of 
quantum noises becomes steady with time, while that under quantum noise effect shown in the main
text could be destroyed after a finite period of time. }}
\vspace{-0.3cm}
\end{figure}

In the general situation the time-ordered exponentials in the solution (\ref{sol}) should be expanded to infinite series (Magnus expansion \cite{expansion}) for numerical calculations. Given a cavity drive with its profile $|E(t)|\leq C$ ($C$ is a constant) such that the function $D(t)$ defined in (\ref{displace}) is bounded, the decay factor $e^{-(\kappa+\gamma_m)(t-\tau)/2}$ dominates the behavior of the matrix $\hat{M}(t,\tau)$, so one has the approximate commutator $[\hat{M}(t,\tau_1),\hat{M}(t,\tau_2)]\approx 0$ in the concerned regimes in which $gE(t)$ is not very large. Then the time-ordered exponentials in the above solution can be replaced by the ordinary exponentials to have a closed form of the solution to the differential equation (\ref{VCM}) as
\begin{eqnarray}
\hat{\vec{v}}(t)&\approx &e^{\int_0^t d\tau \hat{M}(t,\tau)}\hat{\vec{v}}(0)+
\int_0^t  e^{\int_\tau^t d\tau' \hat{M}(t,\tau')}\hat{\vec{f}}(t,\tau)d\tau\nonumber\\
&=& \big(\cosh(\sqrt{m(t,0})\big)\hat{I}+\frac{\sinh\big(\sqrt{m(t,0)}\big)}{\sqrt{m(t,0)}}\hat{K}(t,0)\big)\hat{\vec{v}}(0)\nonumber\\
&+& \int_0^t d\tau\big(\cosh(\sqrt{m(t,\tau})\big)\hat{I}+\frac{\sinh\big(\sqrt{m(t,\tau)}\big)}{\sqrt{m(t,\tau)}}\hat{K}(t,\tau)\big)\hat{\vec{f}}(t,\tau).
\label{result}
\end{eqnarray}
Here we have defined $\hat{K}(t,\tau)=\int_\tau^t d\tau' \hat{M}(t,\tau')$, and the function $m(t,\tau)$ from the relation $\hat{K}^2(t,\tau)=m(t,\tau)\hat{I}$ is
\begin{eqnarray}
m(t,\tau)=\frac{1}{4}|\int_\tau^t d\tau'\big(l_1(t,\tau')+il_2(t,\tau')-il_3(t,\tau')+l_4(t,\tau')\big)|^2-\frac{1}{4}|\int_\tau^t d\tau'\big(l_1(t,\tau')-il_2(t,\tau')-il_3(t,\tau')-l_4(t,\tau')\big)|^2.\nonumber\\
\end{eqnarray}

With arbitrary system parameters, the first term in (\ref{sol}) from the initial
value $\hat{\vec{v}}(0)$ of system operators contributes to one part of the correlation matrix $\hat{V}_1(t)$, where the average in the calculation of the matrix elements is taken with respect to the initial system state $\rho(0)$. This reflects the reliance of the system quantum state at the time $t$ on this initial state. Meanwhile, the second term of noise driving leads to another part of the correlation matrix $\hat{V}_2(t)$, where the average is over the reservoir state $R(0)$. Summing up the two matrices gives the total correlation matrix $\hat{V}(t)=\hat{V}_1(t)+\hat{V}_2(t)$. For a comparison with the entanglement evolution found in the main text,
we give an example of entanglement evolution solely determined by matrix $\hat{V}_1(t)$ in Fig. (C-1).
In the absence of quantum noise effect, the entanglement measured by logarithmic negativity tends to stable value with time, and there does not exist the phenomenon of entanglement sudden death in Fig. 1 and 2 of the main text.
\end{widetext}

\end{document}